\documentclass[12pt]{article}

\usepackage{amsmath,amssymb,amsthm,amscd,pstricks}
\usepackage{graphicx,tocloft}
\usepackage[nosort]{cite}
\def\figurename{Figure}

\makeatletter

\renewcommand{\fnum@figure}[1]{\textbf{\figurename~\thefigure}:}

\makeatother

\makeatletter

\renewcommand\section{\@startsection{section}{1}{\z@}
                                   {-3.5ex \@plus -1ex \@minus -.2ex}
                                   {2.3ex \@plus .2ex}
                                   {\normalfont\large\bfseries}}
\renewcommand\subsection{\@startsection{subsection}{2}{\z@} 
                                   {-3.25ex\@plus -1ex \@minus -.2ex}
                                   {1.5ex \@plus .2ex}
                                   {\normalfont\normalsize\bfseries}}
\renewcommand\subsubsection{\@startsection{subsubsection}{3}{\z@}
                                   {-3.25ex\@plus -1ex \@minus -.2ex}
                                   {1.5ex \@plus .2ex}
                                   {\normalfont\normalsize\bfseries}}
\renewcommand\paragraph{\@startsection{paragraph}{4}{\z@}
                                   {3.25ex \@plus1ex \@minus.2ex}
                                   {-1em}
                                   {\normalfont\normalsize\bfseries}}

\makeatother

\newdimen\tableauside\tableauside=1.0ex
\newdimen\tableaurule\tableaurule=0.4pt
\newdimen\tableaustep
\def\phantomhrule#1{\haox{\vbox to0pt{\hrule height\tableaurule
width#1\vss}}}
\def\phantomvrule#1{\vbox{\haox to0pt{\vrule width\tableaurule
height#1\hss}}}
\def\sqr{\vbox{%
  \phantomhrule\tableaustep

\haox{\phantomvrule\tableaustep\kern\tableaustep\phantomvrule\tableaustep}%
  \haox{\vbox{\phantomhrule\tableauside}\kern-\tableaurule}}}
\def\squares#1{\haox{\count0=#1\noindent\loop\sqr
  \advance\count0 by-1 \ifnum\count0>0\repeat}}
\def\tableau#1{\vcenter{\offinterlineskip
  \tableaustep=\tableauside\advance\tableaustep by-\tableaurule
  \kern\normallineskip\haox
    {\kern\normallineskip\vbox
      {\gettableau#1 0 }%
     \kern\normallineskip\kern\tableaurule}%
  \kern\normallineskip\kern\tableaurule}}
\def\gettableau#1 {\ifnum#1=0\let\next=\null\else
  \squares{#1}\let\next=\gettableau\fi\next}

\newcommand{\be}{\begin{equation}}
\newcommand{\ee}{\end{equation}}
\newcommand{\bea}{\begin{eqnarray}}
\newcommand{\eea}{\end{eqnarray}}
\newcommand{\ba}{\begin{array}}
\newcommand{\ea}{\end{array}}
\newcommand{\id}{\haox{1\kern-.27em l}}
\newcommand{\ZZ}{\mathbb{Z}}

\newcommand{\RR}{\mathbb{R}}
\newcommand{\half}{ {\textstyle \frac{1}{2}  } }
\newcommand{\al}{\alpha}

\newcommand{\ka}{\kappa}

\newcommand{\de}{\delta}

\newcommand{\ep}{\epsilon}

\newcommand{\la}{\lambda}

\newcommand{\De}{\Delta}
\newcommand{\La}{\Lambda}

\newcommand{\tha}{\theta}

\newcommand{\Ups}{\Upsilon}
\newcommand{\cN}{\mathcal{N}}

\newcommand{\cD}{\mathcal{D}}

\newcommand{\cW}{\mathcal{W}}

\newcommand{\D}{{\rm d}}
\newcommand{\pa}{\partial}

\newcommand{\non}{\nonumber}
\newcommand{\lb}{\langle}
\newcommand{\rb}{\rangle}

\newcommand{\SU}{\mathrm{SU}}

\newcommand{\SL}{\mathrm{SL}}

\newcommand{\sll}{\mathrm{sl}}
\newcommand{\U}{\mathrm{U}}

\newcommand{\ts}{\textstyle}

\begin{document}

\begin{center}
\vspace*{2mm}
{\Large\sf
{$\cW$-algebras and surface operators in {\large $\cN=2$} gauge theories }}

\vspace*{6mm}
{\large Niclas Wyllard}

\vspace*{4mm}

{\tt n.wyllard@gmail.com}

\vspace*{8mm}
{\bf Abstract} 
\end{center}
\vspace*{0mm}
\noindent  

A general class of $\cW$-algebras can be constructed from the affine $\sll(N)$ algebra
 by (quantum) Drinfeld-Sokolov reduction and are classified by partitions of $N$. Surface operators in an $\cN=2$ $\SU(N)$ $4d$ gauge theory are also classified by partitions of $N$. 
 We argue that   instanton partition functions of $\cN=2$  gauge theories  in the presence of a surface operator can also be computed from the corresponding $\cW$-algebra. 
 We test this proposal by analysing the Polyakov-Bershadsky $\cW_3^{(2)}$ algebra obtaining results that are in agreement with the known partition functions for $\SU(3)$ gauge theories with a so called simple surface operator.  As a byproduct, our proposal implies relations between the $\cW_3^{(2)}$ and  $\cW_3$ algebras.

\vspace{1mm}

\setcounter{tocdepth}{1}

\setcounter{equation}{0}
\section{Introduction}\label{sint}

In the last year several new detailed connections between $2d$ conformal field theories and $4d$ quiver gauge theories with $\cN\,{=}\,2$ supersymmetry have been discovered. In particular,   conformal (or chiral) blocks \cite{Belavin:1984} of certain $2d$ conformal theories have been argued to be equal to instanton partition functions \cite{Nekrasov:2002} in $4d$ $\cN\,{=}\,2$ quiver gauge theories. 

The starting point of the new developments was the important paper \cite{Alday:2009a} where  a relation  between  the Liouville theory (whose conformal blocks are those of the Virasoro algebra) and instanton partition functions in (conformal) $\cN\,{=}\,2$ $\SU(2)$ quiver gauge theories was uncovered. This result has been extended to various other $2d$ theories, such as the $2d$ $A_{N-1}$ Toda theories, whose conformal blocks are those of the $\cW_N$ algebras, and are conjectured to be related \cite{Wyllard:2009} to instanton partition functions in (conformal) $\cN\,{=}\,2$ $\SU(N)$ quiver gauge theories. Extensions to non-conformal gauge theories have also been discussed, first for $\SU(2)$ theories in \cite{Gaiotto:2009b} and later also for higher rank  theories \cite{Taki:2009}. In addition, conformal blocks of $2d$ conformal field theories with affine $\sll_N$ symmetry have been argued to be related to conformal $\cN\,{=}\,2$ $\SU(N)$ gauge theories in the presence of a so called  full surface operator.  This was first proposed for the  affine $\sll(2)$ conformal blocks in \cite{Alday:2010} and further studied in \cite{Awata:2010}. The extension to $\widehat{\sll}_N$ (affine $\sll_N$) was discussed in \cite{Kozcaz:2010b}.  

In this paper we argue that the above relations are special cases of a general connection between 
 $\cW$-algebras and instanton partition functions in $\cN=2$ gauge theories in the presence of surface operators. 

Before describing our proposal in more detail, we should point out that in parallel to the physics developments there have also been many important  results in  the mathematics literature. For instance, the results in \cite{Carlsson:2008} can be viewed as a simpler version of the AGT relation \cite{Alday:2009a} when the gauge group is $\U(1)$ rather than $\SU(2)$. In the pioneering papers \cite{Braverman:2004a} various aspects of instanton partition functions in the presence of surface operators were discussed. In particular, for the pure $\SU(N)$ theories with a  full surface operator it was shown that the partition function of the gauge theory is equal to the norm of a so called Whittaker vector of the $\widehat{\sll}_N$ algebra. This result can be viewed as a non-conformal version of the AT relation \cite{Alday:2010} and is analogous to the discussion in \cite{Gaiotto:2009b}, which is valid in the absence of surface operators and can also be formulated in the language of  Whittaker vectors (see 
e.g.~\cite{Yanagida:2010}). In a further development \cite{Feigin:2008} explicit expressions for the instanton partition functions of $\SU(N)$ quiver gauge theories in the presence of a full surface operator were determined. Finally, we must also mention the recent paper \cite{Braverman:2010} which contains ideas similar to the ones in this  work, albeit phrased in a more mathematical language.  Phrased in physics terminology, it is shown in  \cite{Braverman:2010}  that the subsector where $4d$ instanton effects decouple of the instanton partition function for the pure $\SU(N)$ theory in the presence of a general surface operator is equal to the norm of a Whittaker vector of a so called finite $\cW$-algebra (a certain finite subalgebra of a $\cW$-algebra). For non-conformal theories, our proposal can be viewed as an extension of the result in \cite{Braverman:2010} to the full $\cW$-algebra (such a possibility was also mentioned in  \cite{Braverman:2010} but was not spelled out explicitly).

A natural class of $\cW$-algebras are obtained  from the $\widehat{\sll}_N$ algebra\footnote{Throughout this paper we focus on the $\widehat{\sll}_N$ algebras and their associated 
$\cW$-algebras, but similar results are expected to hold also for other affine Lie algebras.}
 by quantum Drinfeld-Sokolov reduction (also called hamiltonian reduction).  The $\cW$-algebras that arise from this construction are classified by the embeddings of $\sll_2$ inside $\sll_N$ (or equivalently by the nilpotent orbits or Levi subalgebras of $\sll_N$). Concretely this means that these $\cW$-algebras are classified by partitions of $N$.  The (quantum) Drinfeld-Sokolov reduction method was studied for the $\widehat{\sll}_2$ algebra in \cite{Bershadsky:1989} and shown to lead to the Virasoro algebra upon reduction.  An extension to $\widehat{\sll}_N$ that gives rise to the $\cW_N$ algebras upon reduction was developed in \cite{Feigin:1990} (see also the pioneering work \cite{Fateev:1987}).   In the language of $\sll_2$ embeddings the reductions in \cite{Feigin:1990} correspond to the so called principally embedded $\sll_2$ subalgebras.  The first example of a reduction corresponding to a non-principally embedded $\sll_2$ was obtained in  \cite{Bershadsky:1990} where a reduction from $\widehat{\sll}_3$ gave rise to a previously unknown $\cW$-algebra, now referred to as the Polyakov-Bershadsky  $\cW_3^{(2)}$ algebra  \cite{Polyakov:1989,Bershadsky:1990}. The general connection to $\sll_2$ embeddings  was first observed in the classical case  \cite{Bais:1990} (see also the review \cite{Feher:1992}).  A general theory of quantum reductions for arbitrary $\sll_2$ embeddings was developed in \cite{deBoer:1993} (see also e.g.~\cite{Kac:2003a} for some further mathematical developments.)

 One way to define a surface operator in a $4d$ gauge theory is by specifying the (singular) behaviour of the gauge field (and scalars, if present) near the $2d$ submanifold where the surface operator is supported. In \cite{Gukov:2006} it was found that the possible types of surface operators in an $\cN\,{=}\,4$ $\SU(N)$ gauge theory are in one-to-one correspondence with the Levi subalgebras of $\SU(N)$.  Concretely this means that for every (non-trivial) partition of $N$ there is a possible surface operator. Surface operators in $4d$ $\SU(N)$ theories with $\cN\,{=}\,2$ supersymmetry are also classified by partitions of $N$ and have been studied e.g.~in \cite{Gukov:2007} (and more recently in the context of the AGT relation in several papers \cite{Alday:2009b,Gaiotto:2009c,Kozcaz:2010,Dimofte:2010,Maruyoshi:2010,Taki:2010,Alday:2010,Awata:2010,Kozcaz:2010b}).  For $\cN\,{=}\,2$ theories a surface operator depends on  a certain number of continuous complex parameters, one for each abelian $\U(1)$ factor in the Levi subalgebra. Following \cite{Alday:2010} we call   a surface operator corresponding to the partition $N=(N{-}1)+1$ a simple surface operator and a surface operator corresponding to $N=1{+}\ldots{+}1$ a full surface operator. 

As was recalled above, both the $\cW$-algebras that are obtained by quantum  Drinfeld-Sokolov reduction from the $\widehat{\sll}_N$ algebra and  surface operators in  $\cN=2$ $\SU(N)$ gauge theories are classified by partitions of $N$. We argue that this is not a coincidence and that the two classes of objects are related.

 We propose that  instanton partition functions of $\cN\,{=}\,2$ $\SU(N)$ gauge theories in the presence of a surface operator corresponding to a given partition of $N$  are also computable from  the $\cW$-algebra corresponding to the same partition. For non-conformal gauge theories the relevant $\cW$-algebra quantity is the norm of a Whittaker vector, whereas for conformal gauge theories the relevant object is a conformal block. This proposal generalises in a very natural way the two cases previously considered in the literature: Whittaker vectors/conformal blocks of the $\widehat{\sll}(N)$ algebra have been shown/argued \cite{Braverman:2004a,Alday:2010,Kozcaz:2010b} to correspond to non-conformal/conformal $\SU(N)$  instanton partition functions with a full surface operator  and conformal blocks/Gaiotto states of the $\cW_N$ algebras correspond \cite{Alday:2009a,Wyllard:2009, Gaiotto:2009b,Taki:2009} to conformal/non-conformal $\SU(N)$ instanton partition functions in the absence of a surface operator.  In the language of partitions, these two cases  correspond to the partitions $N=1{+}\cdots{+}1$ and $N=N$, respectively. 
 
 In the next section we test our proposal by analysing the Polyakov-Bershadsky $\cW_3^{(2)}$ algebra. This $\cW$-algebra corresponds to the partition $3=2{+}1$ and is the simplest case which has not previously been studied.  
Our proposal implies that it should be possible to use $\cW_3^{(2)}$ methods to compute partition functions in $\cN\,{=}\,2$ $\SU(3)$ gauge theories with a simple surface operator. Such partition functions have previously been computed using other approaches\footnote{Strictly speaking the simple surface operator appearing in these papers, although also associated with $3=2+1$, is not precisely the same as the one that appears in the $\cW_3^{(2)}$ computation. However our results (as well as those in \cite{Awata:2010,Kozcaz:2010}) indicate that for the purpose of computing the instanton partition function they can be considered to be the same (at least for non-quiver theories). In this paper both types will therefore be referred to as a simple surface operator (see section \ref{sdisc} for a further discussion).} \cite{Alday:2009a,Kozcaz:2010,Dimofte:2010,Taki:2010}. Using these results we find agreement with $\cW_3^{(2)}$ computations. As a byproduct we find relations between the $\cW_3^{(2)}$ and $\cW_3$ algebras.

\setcounter{equation}{0} 
\section{$\cW$-algebras and surface operators for rank two } \label{A2}

In this section we test the idea outlined above relating $\cW$-algebras and instanton partition functions  in $\cN=2$ gauge theories with surface operators. We focus on  the rank two theories. For such theories, the partition $3=1{+}1{+}1$ corresponds to the $\widehat{\sll}(3)$ algebra (no reduction) and to a full surface operator in $\cN \,{=}\,2$ $\SU(3)$ gauge theories,  the partition $3=3$ corresponds to the reduction of $\widehat{\sll}(3)$ to the $\cW_3$ algebra \cite{Zamolodchikov:1985} and to the absence of a surface operator. The final case, $3=2{+}1$, corresponds to the reduction of  $\widehat{\sll}(3)$ to the $\cW^{(2)}_3$ algebra \cite{Polyakov:1989,Bershadsky:1990} and to a simple surface operator. We summarise the various possibilities in the following table:

\medskip

\begin{center}
\begin{tabular}{|c|c|c|}
\hline
Partition & $2d$ symmetry algebra& Type of surface operator\\
\hline
$\phantom{\bigg(} \!\!\! 1{+}1{+}1$& $\widehat{\sll}(3)$& Full\\
\hline
$\phantom{\bigg(} \!\!\! 2{+}1$& $\cW_3^{(2)}$ & Simple\\
\hline
$\phantom{\bigg(} \!\!\! 3$& $\cW_3$& Absent \\ \hline
\end{tabular}
\end{center}

\medskip

The relation between the second and third columns in the last row is the $A_2$ AGT relation \cite{Wyllard:2009,Alday:2009a} (or its non-conformal version \cite{Taki:2009,Gaiotto:2009b}) and in the first row the $A_2$ AT relation \cite{Kozcaz:2010b,Alday:2010} (or its non-conformal version \cite{Braverman:2004a}). The relation between the last two columns for the middle row is the subject of this section and constitutes the first previously unknown case illustrating our proposal relating $\cW$-algebras and surface operators in $\cN=2$ gauge theories.

 We first review various properties of the $\cW_3^{(2)}$ algebra and its representations and then in section \ref{sW32pert} perform some perturbative computations. These results should be compared to  instanton partition functions in $\SU(3)$ theories with a simple surface operator. In the general case we do not know how to compute the instanton partition function in the presence of a surface operator. However, for the case of a simple surface operator one can fortunately use the alternative dual description in terms of a degeneratate field in the $\cW_3$ algebra ($A_2$ Toda theory)  \cite{Alday:2009b,Drukker:2010,Kozcaz:2010,Dimofte:2010}. Using this result, in section \ref{sdeg} we perform some perturbative $\cW_3$ computations (with a degenerate field insertion),  finding complete agreement with the  $\cW_3^{(2)}$ computations in section \ref{sW32pert}.

\subsection{The $\cW^{(2)}_3$ algebra  and its representations } \label{sW32}

The Polyakov-Bershadsky $\cW_3^{(2)}$ algebra \cite{Polyakov:1989,Bershadsky:1990} is an extension of the Virasoro algebra. In addition to the energy-momentum tensor $T(z)$ it also contains two fields $G^{\pm}(z)$ each with conformal dimension $3/2$ and one field $J(z)$ with conformal dimension $1$. These fields have the mode expansions
\be \label{modes}
J(z) = \sum_n z^{-n-1} J_n \,, \qquad G^{\pm}(z) = \sum_n z^{-n-\frac{3}{2}} G^{\pm}_n \,, \qquad T(z) = \sum_n z^{-n-2} L_n \,.
\ee
The modes satisfy the following commutations relations (which are straightforwardly obtained from the more commonly quoted operator product expansions) 
\bea \label{W32}
&& \!\!\!\!\!  \!\!\!\!\! [L_n,J_m] = -m\, J_{n+m}\,, \qquad [L_n,G_m^{\pm}] = (\frac{n}{2}{-}m)\, G^{\pm}_{n+m}\,, \qquad [J_n,G_m^{\pm}] = \pm G_{n+m}^{\pm} \,, \non \\
&&   \!\!\!\!\!  \!\!\!\!\!  {}[J_n,J_m] = \frac{2k+3}{3} \,n \, \de_{n+m,0} \,, \quad \; [L_n,L_m] = (n{-}m)L_{n+m} + \frac{c}{12} n(n^2-1)   \de_{n+m,0}  \,, \\ 
&&  \!\!\!\!\!  \!\!\!\!\!  {}[G^+_n,G^-_m] = \frac{(k+1)(2k+3)}{2}(n^2{-}{\ts \frac{1}{4} }) \de_{n+m,0} - (k{+}3) L_{n+m}  +\frac{3}{2} (k{+}1) (n{-}m) J_{n+m} \non \\ 
&&  \!\!\!\!\!  \!\!\!\!\! \qquad \qquad\, +\, 3  \,\sum_\ell : J_{n+m-\ell}J_\ell:   \non 
\eea
where $k$ is a parameter,  $c=-\frac{(2k+3)(1+3k)}{k+3}$ and  $:\;:$ denotes the normal ordering 
\be
: X_n Y_m: = \left\{ \ba{c}   X_n Y_m \qquad {\rm if} \qquad n \le  m \\     
Y_m X_n \qquad {\rm if} \qquad n > m\ea \right.
\ee

The $\cW^{(2)}_3$ algebra is similar to the well-known $\cN\,{=}\,2$ superconformal algebra \cite{Ademollo:1975}, but in (\ref{W32}) $G^{\pm}_n$ are bosonic and there is a nonlinear $J^2$ term in the algebra. Despite these differences it is still true that one can consider both Ramond and Neveu-Schwarz sectors. These differ by whether $n$ in the mode-expansion of $G^{\pm}(z)$ in (\ref{modes}) are integers or half-integers. 

We mainly consider the Ramond sector, where $G^{\pm}_n$ are integer moded. 
The zero-mode sector of (\ref{W32}) is of particular importance and is spanned by  $J_0$, $G^{\pm}_0$, and $L_0$. Introducing the notation 
\be \label{finitegens}
H = 2J_0 \,, \qquad E = 2G_0^+ \,, \qquad F=  \frac{2}{3}G^-_0  \,, \qquad C= -\frac{4(k{+}3)}{3}L_0 -\frac{(k{+}1)(2k{+}3)}{6}\,,
\ee
we find the algebra 
\be \label{finiteW}
[H,E]= 2E \,, \qquad [H,F]= -2F \,, \qquad [E,F] = H^2 + C\,.
\ee
This is an example of a so called finite $\cW$-algebra \cite{Tjin:1992}.  Finite $\cW$-algebras can be obtained by (quantum) Drinfeld-Sokolov reduction from {\it ordinary} Lie algebras (rather than from affine Lie algebras) \cite{Tjin:1992,deBoer:1992}. The above algebra (\ref{finiteW}) arises via reduction  from $\sll_3$  \cite{Tjin:1992,deBoer:1992}. (See \cite{DeSole:2005} for a discussion of various equivalent ways of defining a finite $\cW$-algebra and their relations to $\cW$-algebras.)  As discussed in \cite{DeSole:2005} it is the Ramond sector that is most directly related to the finite $\cW$-algebra.

The representation theory for the $\cW^{(2)}_3$ algebra has been developed in the literature. 
In the Ramond sector, a highest weight (or primary) state $|\la \rb$ satisfies \cite{Kac:2004}
\be \label{eigenvals}
 L_0 |\la \rb =  \left( \frac{ \lb \la, \la - (k+1)\rho \rb}{2(k+3)} -\frac{1}{8} \right) |\la \rb \,, \qquad  J_0 |\la \rb =  \left( \lb \la,h_2\rb -\half \right) |\la \rb  \,,
 \ee
 together with
 \be \label{anni}
L_n |\la \rb =  G^{+}_{n-1} |\la \rb =  G^{-}_{n} |\la \rb =  J_n |\la \rb =  0  \qquad (n=1,2,\ldots)\,.
\ee

In (\ref{eigenvals})  $\la$ denotes a vector in the root/weight space of $\sll_3$, i.e.~$\la = \la^1 \La_1 + \la^2 \La_2$ where $\La_{1,2}$ are the two fundamental weights of $\sll_3$. Furthermore, $\rho=\La_1+\La_2$ is the Weyl vector and $h_2 = \La_2-\La_1$ (see appendix \ref{ALie} for more details of our Lie algebra conventions).  Note that shifting $\la$ in (\ref{eigenvals}) by a term proportional to $\rho$  changes the form of the $L_0$ eigenvalue, but does not change the $J_0$ eigenvalue. The representation theory in the Ramond sector is closely related to the representation theory of the associated finite $\cW$-algebra (\ref{finiteW}). The representation theory of the algebra (\ref{finiteW}) was obtained in \cite{Tjin:1992,deBoer:1992} (see also \cite{Smith:1990}).

The Neveu-Schwarz version of (\ref{eigenvals}), (\ref{anni}) can be found e.g.~in \cite{Furlan:1994}. In this case the $\la$-independent terms in (\ref{eigenvals}) are absent and the $G^{\pm}$ conditions in (\ref{anni}) are replaced by $G^{\pm}_r |\la\rb = 0$ for all positive half-integers $r$.

In the Ramond sector, the descendants of a primary  state, $\lb \la |$, are  denoted $\lb {\bf n} ;\la |$, where  
\be
\lb {\bf n}; \la | = \lb \la | G^{+}_{n^+_1-1} \cdots G^{+}_{n^+_{\ell^+}-1}  G^{-}_{n^-_1} \cdots G^{-}_{n^-_{\ell^-}} J_{n_1} \cdots J_{n_\ell} \,  L_{\tilde{n}_1} \cdots L_{\tilde{n}_\ell}  \,,
\ee
and $n^\pm_i$, $\tilde{n}_i$  and $n_i$ can be any positive integer.  Similarly,
 \be
| {\bf n}; \la \rb  =  L_{-\tilde{n}_1} \cdots L_{-\tilde{n}_\ell} \, J_{-n_1} \cdots J_{-n_\ell} \, G^{+}_{-n^+_1} \cdots G^{+}_{-n^+_{\ell^+}}  G^{-}_{-n^-_1+1} \cdots G^{-}_{-n^-_{\ell^-}+1}   |\la \rb  \,.
\ee
The matrix of inner products of descendants (usually called the  Gram or Shapovalov matrix) satisfies 
\be  \label{Xmatrix}
X_{ \la}( {\bf n} ; {\bf m })  = \lb {\bf n};\la |  {\bf m}; \la \rb\propto \delta_{ \rm N , \rm M}\,  \delta_{ S_{n}+S_{m},0} \,,
\ee
i.e.~it is a block-diagonal matrix where each block contains only  descendants with  given values for the total level ${\rm N}= \sum_i (n_i + \tilde{n}_i + n^{+}_i+ n^{-}_i) $ and the total charge, $S_n$, given by the number of $n^{+}_i$ minus the number of $n^{-}_i$.

\subsection{Perturbative computations for the $\cW^{(2)}_3$ algebra} \label{sW32pert}

A Whittaker-type state (vector) can be defined for the $\cW^{(2)}_3$ algebra in a way completely analogous to the construction in \cite{Braverman:2004a,Braverman:2010} (see also section 5 in \cite{Kozcaz:2010b} for a discussion using the notation of \cite{Gaiotto:2009b} that will also be used below).  We denote this state by $|x_1,x_2; \la \rb$ and demand that it should satisfy
\be \label{Wconds}
G_0^{+} |x_1,x_2; \la \rb = \sqrt{x_1} \,|x_1,x_2 ; \la \rb \,, \qquad G_1^{-} |x_1,x_2; \la\rb = \sqrt{x_2}\, |x_1 ,x_2; \la \rb \,, 
\ee
where all other $G^{\pm}_n$, $J_n$ and $L_n$ that annihilate $| \la \rb$ also annihilate $|x_1,x_2;\la \rb$. The norm of the Whittaker state can be expressed in terms of certain (diagonal) components of the inverse of the matrix (\ref{Xmatrix}). The following set of descendants play a distinguished role in this construction
\be 
|n,p; \la \rb = (G_{-1}^{+})^p  (G_{0}^{-})^n |\la\rb \,.
\ee
Denoting the corresponding diagonal component of the inverse of the matrix $X_{\la}$ by $X^{-1}_\la(n,p;n,p)$, the norm of the Whittaker vector can be obtained via
\be \label{whit}
\lb x_1,x_2;\la |x_1,x_2 ; \la \rb = \sum_{n=0}^{\infty} \sum_{p=0}^{\infty} X_\la^{-1}(n,p;n,p) \, x_1^n \, x_2^p\,.
\ee
 From our proposal it follows that this expression should equal (possibly up  to a prefactor)  the instanton partition  function  for  the pure $\cN\,{=}\,2$ $\SU(3)$ theory with a simple surface  operator insertion. 

The terms in (\ref{whit}) containing only $x_1$ involve descendants of the form $(G_0^-)^n | \la \rb$. For such descendants, the  Gram matrix  is  diagonal and can be computed using (\ref{W32}), (\ref{eigenvals}) and (\ref{anni}) with the result  
\bea
\lb \la|(G_0^+)^n (G_0^-)^n |\la \rb &=& n (\la_1 {-} {\ts \frac{k}{2}} {+}{\ts \frac{1}{2}} {+}n{-}1)(-\la_2 {+} {\ts \frac{k}{2}} {+}{\ts \frac{3}{2}} {+} n{-}1)\lb \la | (G_0^+)^{n-1} (G_0^-)^{n-1}|\la  \rb \non \\ &=& n! \, (\la_1 -{\ts \frac{k}{2}} +{\ts \frac{1}{2}})_n (-\la_2 + {\ts \frac{k}{2}} +{\ts \frac{3}{2}})_n \,,
\eea
where $(X)_n= X (X+1) \cdots (X+n-1)$ is the usual Pochhammer symbol. The contribution to (\ref{whit}) is consequently   
\be\label{x1}
 \sum_{n=0}^{\infty}\frac{  1  }{   (\la_1 -{\ts \frac{k}{2}} +{\ts \frac{1}{2}})_n (-\la_2 + {\ts \frac{k}{2}} +{\ts \frac{3}{2}})_n } \frac{x_1^n}{n!}\,.
\ee
Similarly, the terms depending only on $x_2$ arise from the result  
\be
\lb \la|(G_1^-)^n (G_{-1}^+)^n |\la \rb = (-1)^n n! \, (-\la_1 -{\ts \frac{k}{2}} -{\ts \frac{3}{2}})_n (\la_2 - {\ts \frac{3k}{2}} -{\ts \frac{5}{2}})_n \,,
\ee
and lead to the following contribution to (\ref{whit}) 
\be\label{x2}
 \sum_{n=0}^{\infty}\frac{  1  }{   (-\la_1-{\ts \frac{k}{2}} - {\ts \frac{3}{2}})_n (\la_2 - {\ts \frac{3k}{2}} - {\ts \frac{5}{2}})_n  } \frac{(-x_2)^n}{n!}\,.
\ee
It is also possible to compute subleading terms. As an example, we consider the terms of the form $x_1^{n+1} \,  x_2$. The relevant block of the Gram matrix involve descendants of the form 
\bea
&& |1\rb=  G_{-1}^{+} (G_0^{-} )^{n+1}  | \la \rb \,,   \qquad |2\rb=G_{-1}^{-} (G_0^{-} )^{n-1}  | \la \rb \,, \non \\
&& |3\rb=J_{-1} (G_0^{-} )^n  | \la \rb \,,  \qquad \quad \,   |4\rb=L_{-1} (G_0^{-} )^n  | \la\rb \,.
\eea
For  any $n\geq 1$ these  states  generate  a $4{\times}4$  sub-block  $X_{r,s}= \lb r | s\rb$ with $r,s=1,\ldots,4$ of the  Gram matrix:\footnote{When $n=0$,  the block  reduces  to a $3{\times}3$  block (obtained from (\ref{G44}) by removing the 2nd row and column and setting $n=0$).} 
 \be \label{G44}
X_{r,s}= \left( \ba{cccc} P_1(\la) M(n{+}1)& 0 &M(n{+}1) & {\ts \frac{3}{2}} M(n{+}1)  \\
 0& P_2(\la) M(n{-}1) & -M(n) & {\ts \frac{3}{2}}M(n) \\ 
 M(n{+}1)  & -M(n)&{\ts \frac{(2k+3)}{3}}M(n) & [q(\la)-n] M(n)  \\
  {\ts \frac{3}{2}} M(n{+}1) & {\ts \frac{3}{2}}M(n) &[q(\la)-n]  M(n) &2 \De(\la)M(n) \ea \right)
\ee
with
\bea
\!\!\!\!\!\!P_1(\la)&\!\! =\!\! & -\frac{3(k{+}1)(2k{+}3)}{8} +(k{+}3)\De(\la) +3(k{+}1)[\Ups(\la){-}n{-}1]-3[\Ups(\la){-}n{-}1]^2 \,, \non  \\
\!\!\!\!\!\!P_2(\la) &\!\! =\!\! & \frac{3(k{+}1)(2k{+}3)}{8} -(k{+}3)\De(\la) +3(k{+}1)[\Ups(\la){-}n{+}1]+3[\Ups(\la){-}n{+}1]^2 \,,
\eea
where $\De(\la)$ denotes the eigenvalue of $L_0$ in (\ref{eigenvals}), $\Ups(\la)$ denotes the $J_0$ eigenvalue, and
\be
M(n)\equiv {\lb \la |  (G_0^{+} )^{n}     (G_0^{-} )^{n}   | \la \rb }= n! \, (\la_1 -{\ts \frac{k}{2}} +{\ts \frac{1}{2}})_n (-\la_2 + {\ts \frac{k}{2}} +{\ts \frac{3}{2}})_n \,.
\ee
Inverting (\ref{G44}) and selecting the 1,1 component in accordance with the general result  (\ref{whit}), gives a closed expression for all $x_1^{n+1} \,  x_2$ terms. However,  as this expression is  somewhat unwieldy we only give the coefficient of the $x_1 x_2$ term:
\be \label{x1x2}
\frac{ 8 (9 {+} 6 k {+} k^2 {+} 12[1 {+} k] \la_1 {+} 4 k^2 \la_1 {-} 2 [1{+} k] \la_1^2 {+} 8 k \la_2 {+} 4 k^2 \la_2 {-} 8 \la_1 \la_2 {-} 4 k \la_1 \la_2 {-} 
   2 [1{+}  k] \la_2^2)}
   {  (k{+}3) (k{-}1 {-} 2 \la_1) (k{+}3  {+}  2 \la_1) (k{+}3 {-} 2 \la_2) (3k{+}5 {-} 2 \la_2) (2k{+}3 {-} \la_1 {-} \la_2) (1 {+}
    \la_1 {+} \la_2)}.
  \ee
 So far we have focused on $\cW_3^{(2)}$ quantities that on the gauge theory side correspond to the (non-conformal) pure $\SU(3)$ theory. It should also be possible to consider conformal  $\SU(3)$ gauge theories. For instance, from our proposal and standard AGT-type  arguments it follows that the four-point $\cW_3^{(2)}$ conformal block on the sphere  should equal (possibly up  to a prefactor)  the instanton partition  function  for  the $\cN\,{=}\,2$ $\SU(3)$ theory with $N_f=6$ and a simple surface  operator insertion. It seems natural to assume that the primary field  corresponding to the state $|\la\rb$ can be expressed as $V_{\la}(x,z)$, where $x$ is an isospin variable and $z$ denotes the worldsheet coordinate. In the standard decomposition, the four-point conformal block can then be written
\be\label{4pt}
 \!\!\! \!\!\! \!\!\!   \sum_{{\bf n}; {\bf p} }
\frac{   \lb \la_1| V_{\xi_2}(1,1)  |{\bf n};  \la\rb  
 X^{-1}_{\la}( {\bf n} ; {\bf m } ) 
  \lb {\bf m }; \la |  V_{\xi_3}(x,z) |\la_4 \rb }{      \lb \la_1| V_{\xi_2}(1,1)  | \la \rb \lb \la | V_{\xi_3}(x,z) |\la_4 \rb    } \,.
\ee
As in \cite{Wyllard:2009,Kozcaz:2010b} the $\xi_i$ should be special (restricted) momenta which  should lead to crucial simplifications. To compute (\ref{4pt}) one would in particular need to know the commutation relations between the generators of the  $\cW_3^{(2)}$ algebra and the $V_{\xi_i}$'s.  As in the $\widehat{\sll}_3$ case it is natural to expect that  these commutation relations can be expressed in terms of differential operators acting on the isospin (and worldsheet) variables (as in the $\cW_3$ case there can also be pieces that can not be expressed as differential operators). One encouraging result is that the zero-mode part of the $\cW_3^{(2)}$  algebra (i.e.~the finite $\cW_3^{(2)}$-algebra) can be realised in terms of differential operators as (see also the discussion in section 6 of  \cite{deBoer:1992}) 
\bea \label{Dzs}
&& D_0^+ =  -x \left[ \frac{(k{+}1)(2k{+}3)}{8} + (k{+}3)x [\De + z\pa_z] - 3 \Ups^2\right] - x^2(3\Ups-{\ts \frac{3}{2}} )\frac{\D}{\D x}  + x^3 \frac{\D^2}{\D x^2}   \,, \non   \\
&&  
 D_0^- =   \frac{\D}{\D x}   \,, \qquad  \quad D_0 =     \Ups  - x  \frac{\D}{\D x}   \,, \qquad \quad
\cD_0 = \De + z \pa_z \,,
\eea
where $\De$ denotes the conformal dimension (the eigenvalue of $L_0$ in (\ref{eigenvals})), and $\Ups$ denotes the $J_0$ eigenvalue. (Note that the algebra (\ref{Dzs}) also closes if one omits the $z\pa_z$ terms.)

We should also mention that in the $\widehat{\sll}_N$ computations in  \cite{Alday:2010,Kozcaz:2010b} additional operator  insertions in the conformal blocks were crucial to obtain agreement with the instanton computations. Similar insertions are probably also required in the $\cW_3^{(2)}$ case. 

As there are several unsolved (technical) problems associated with the computations of conformal blocks for the  $\cW_3^{(2)}$ algebra we postpone a full discussion to future work.

\subsection{ $\cW_3$  degenerate fields and simple surface operators } \label{sdeg}

Instanton partition functions for $\cN\,{=}\,2$ $\SU(3)$ gauge theories can be obtained from the $\cW_3$ algebra \cite{Wyllard:2009,Taki:2009} (see also \cite{Mironov:2009a,Kanno:2009}). The addition of a certain simple surface operator can be interpreted as the insertion of a degenerate field in the $2d$ CFT  \cite{Alday:2009b,Drukker:2010,Kozcaz:2010,Dimofte:2010}.  For the pure $\SU(3)$ theory the relevant quantity is 
\be \label{W3whit}
\lb  y ;\al |V_{-b\La_1}(x) |y ; \al \rb ,
\ee
where (in our notation) $ |y ; \al \rb$ is the $\cW_3$ (Whittaker) state constructed in \cite{Taki:2009} and $V_{-b\La_1}$ is a degenerate field of the $\cW_3$ algebra. For the conformal $\SU(3)$ theory with $N_f=6$ the relevant quantity is a  particular five-point  $\cW_3$ conformal  block  where two of the insertions are special (cf.~\cite{Wyllard:2009}) and one  of  the  insertions  is  the  degenerate field $V_{-b\La_1}$.  

An alternative to the $\cW_3$ degenerate field approach is to use the (B or A model) topological string description of a simple surface operator \cite{Kozcaz:2010,Dimofte:2010}, or the gauge theory method in  \cite{Kozcaz:2010} which uses a combination of  the conjectures in \cite{Alday:2009a} and \cite{Alday:2009b}  and corresponds to a geometric transition in the topological string language \cite{Dimofte:2010,Taki:2010}.  

We first briefly describe the $\cW_3$ approach. Primary fields associated with $\cW_3$ are denoted $V_\al(z)$ where $\al = \al^1 \La_1 + \al^2 \La_2$, and the corresponding state is denoted $|\al\rb$. By inserting two complete  sets  of  states  the five-point $\cW_3$ conformal block mentioned above can be written (we suppress the three-point factors in the denominator) 
\bea \label{4+1}
\!\! \!\! \sum_{{\bf n},{\bf n}',{\bf m},{\bf m}'} \!\! \lb \al_1| V_{\chi_2}(1) |{\bf n};\al \rb X^{-1}_{ \bf n; \bf n'}(\al)
\lb {\bf n}';\al | V_{-b\La_1}(x)  |{\bf m};\tilde{\al} \rb X^{-1}_{ \bf m; \bf m'}(\tilde{\al })
\lb {\bf m}';\tilde{\al } |  V_{\chi_3}(z) |\al_4 \rb , 
\eea 
where  $\chi_i = \ka_i \La_1$,  $|{\bf n}; \al \rb$ is short-hand notation for the descendants of the primary state $|\al \rb$,  $X^{-1}_{ \bf n; \bf n'}(\al )$  is  the  inverse of  the Gram matrix, and  the  sums  run over are  all  descendants.  The  terms in (\ref{4+1}) with $|{\bf m}; \al \rb  =|{\bf m'}; \al \rb =| \al \rb$ depend only on $x$ and  after summing over ${\bf n}$ and ${\bf n}'$ reduce to 
 \bea \label{xterms}
 \lb \al_1| V_{\chi_2}(1) V_{-b \La_1 }(x) |\tilde{\al }\rb \propto \,{}_3 F_2(A_1,A_2,A_3;B_1,B_2;x)\,,
  \eea
where we used the results in  \cite{Fateev:2005}. This result has also been obtained from the dual gauge theory  \cite{Mironov:2009a} and was discussed in  \cite{Schiappa:2009} using the matrix model approach \cite{Dijkgraaf:2009} (see also \cite{Cheng:2010}). The hypergeometric function in (\ref{xterms})  is defined in the neighbourhood  of  $x=0$ and has the series expansion 
\be \label{3F2}
{}_3F_2(A_1,A_2,A_3;B_1,B_2;x) = \sum_{n=0}^{\infty} \frac{(A_1)_n (A_2)_n(A_3)_n}{(B_1)_n(B_2)_n}\frac{x^n}{n!} \,,
\ee
with   
\be
A_i = b( {\ts \frac{1}{3}}\ka_2  - {\ts {\frac{2}{3} }b + \lb \tilde{\al } {-}Q \rho ,h_1}\rb - \lb \al_1{-} Q \rho ,h_i\rb) \,, \quad B_i = 1+b\lb \tilde{\al }{-}Q\rho,h_1{-}h_{i+1}\rb \,.
\ee
Similarly, the terms with $|{\bf n}; \al \rb  =|{\bf n'}; \al \rb =| \al \rb $ depend only on $\frac{z}{x}$ and  reduce to 
\be \label{2ndcase}
\!\! \!\! \lb \al  | V_{-b \La_1 }(x) V_{\chi_3}(z) |\al_4 \rb =\propto {}_3 F_2(C_1,C_2,C_3;D_1,D_2 ;\frac{z}{x}) \,,
\ee
where 
\be \label{cd}
C_i = b( {\ts \frac{1}{3}}\ka_3  - {\ts {\frac{2}{3} }b + \lb \al_4 {-}Q \rho ,h_i}\rb - \lb \al {-} Q \rho ,h_1\rb) \,, \quad D_i = 1-b\lb \al{-}Q\rho,h_1{-}h_{i+1}\rb\,.
\ee
The above expressions correspond on the gauge theory side to the conformal $\SU(3)$ theory with $N_f=6$; the expressions relevant to the pure $\SU(3)$ theory can be obtained by taking the``non-conformal limit" i.e.~by replacing the $(A_i)_n$ and $(C_i)_n$ factors by 1. Alternatively, one can analyse (\ref{W3whit}) directly using the method in \cite{Awata:2009,Awata:2010}. 

By comparing the non-conformal version of the above two results to the corresponding results in the previous subsection (\ref{x1}), (\ref{x2}) we see that they agree provided we make the identifications
\be
x_1 = x\,, \quad x_2 = -\frac{z}{x} \,,\quad  k{+}3 = -b^2\,, \quad \la_1 = b \,\al_1 -\frac{b^2}{2} -2\,,\quad \la_2 = -b(\al_1{+}\al_2)+\frac{b^2}{2} +1\,. 
\ee
Furthermore, $\tilde{\al }=\al + b\La_1$, which is simply 
the degenerate fusion rule. 

We have also analysed a class of subleading terms.  These can be obtained from CFT considerations as above, but we found it more convenient to use the method in section 6 of \cite{Kozcaz:2010}. In this method the partition function of the $\SU(3)$ gauge theory with a simple surface operator is obtained from an  $\SU(3){\times}\SU(3)$ quiver gauge theory (with instanton expansion parameters $y_1$ and $y_2$) by imposing certain restrictions, which are simply the degenerate field and fusion requirements translated into gauge theory language using the AGT relation. 
Using this method the coefficient in front of the $y_1y_2$ term in the instanton partition function for the pure $\SU(3)$ theory with a simple surface operator becomes (here $\ep \equiv\ep_1+\ep_2$)
\be \label{y1y2}
\frac{(-6 a_1^2 \epsilon_1 {-} 6 a_1 a_2 \epsilon_1 {-} 6 a_2^2 \epsilon_1 {+} 6 \epsilon_1^3 {-} a_1^2 \epsilon_2{ -}  4 a_1 a_2 \epsilon_2 {-} 4 a_2^2 \epsilon_2 {+} 3 a_1 \epsilon_1 \epsilon_2 {+} 10 \epsilon_1^2 \epsilon_2 {+} 
 5 \epsilon_1 \epsilon_2^2 {+} \epsilon_2^3)}
 {\epsilon_1^2 \epsilon_2 (  \epsilon_1{+}a_1 {-} a_2) (\epsilon_1 {+}2 a_1 {+} a_2 ) (\epsilon{-}a_1 {-}  2 a_2) ( \epsilon {-}2 a_1 {-} a_2) ( \epsilon{-}a_1 {+} a_2 ) (\epsilon{+}a_1 {+}  2 a_2 )},
\ee
where $a_{1,2}$ are the $\SU(3)$ Coulomb parameters. The result (\ref{y1y2}) matches (\ref{x1x2}) provided that
\be
\! x_1 = y_1\,, \quad x_2 = -y_2 \,, \quad    k{+}3 = -\frac{\ep_2}{\ep_1} \,, \quad \la_1 = \frac{a_2{-}a_1}{\ep_1} +\frac{1}{2} \frac{\ep_2}{\ep_1}-1 \,, \quad \la_2 = \frac{2a_1{+} a_2}{\ep_1} -\frac{3}{2} \frac{\ep_2}{\ep_1}-1\,.
\ee
The leading  $y_1^n$  and $y_2^n$ terms of course also match, as do higher-order  $y_1^n y_2$ terms. The non-trivial agreement of these infinite sets of terms supports our idea that  instanton partition functions in $\cN\,{=}\,2$ $\SU(3)$ gauge theories with a simple surface operator should be computable from the $\cW_3^{(2)}$ algebra. 

As a byproduct of our analysis we find relations between the $\cW_3^{(2)}$ and $\cW_3$ algebras. For the non-conformal case the conjecture is that (\ref{whit}) is equal to (\ref{W3whit});   more generally there should also be relations between $\cW_3^{(2)}$ conformal blocks  and $\cW_3$ conformal blocks  with an additional degenerate field insertion, e.g.~we expect that (\ref{4pt}) and (\ref{4+1}) should be equal (possibly up to a prefactor).

\setcounter{equation}{0} 
\section{Discussion} \label{sdisc}

In this paper we argued that there is a general connection between $\cW$-algebras and instanton partition functions in $\cN\,{=}\,2$ gauge theories with surface operators (similar ideas were  discussed in \cite{Braverman:2010}). This proposal is very natural from the viewpoint in \cite{Gaiotto:2009a} which uses the $6d$ $(2,0)$ theory formulated on $\RR^4{\times}C$, where an $\cN\,{=}\,2$ $\SU(N)$ gauge theory lives on $\RR^4$ and the $2d$ conformal field theory lives on the Riemann surface $C$.  As discussed in \cite{Alday:2010}, one way a surface operator can  arise is from a $4d$ defect spanning a $2d$ submanifold of $\RR^4$ and wrapping $C$. In \cite{Gaiotto:2009a} it was argued that for the $\SU(N)$ theories, the $4d$ defects of the $(2,0)$ theory are classified by Young tableaux or equivalently by partitions of $N$, so the class of surface operators constructed from $4d$ defects should also be classified by partitions. Thus in this construction it should be possible to describe  a general surface operator. Our proposal can be viewed as a prescription for how the symmetry algebra of the $2d$ theory is changed when a general $4d$ defect wraps $C$. 

It is also possible to describe surface operators using $2d$ defects spanning a submanifold inside $\RR^4$ and intersecting $C$ at a point. This construction leads to the interpretation of a simple surface operator, i.e.~a surface operator corresponding to the partition $N=(N{-}1)+1$, in terms of degenerate fields in the $A_{N-1}$ Toda theory as first proposed in in \cite{Alday:2009a}. It is less clear (at least to us) if one can describe general surface operators using only $2d$ defects. But at least for a simple surface operator there are two descriptions, in terms of $4d$ or $2d$ defects.  There are certain differences between the simple surface operators that arise from these two constructions, but computations in \cite{Awata:2010,Kozcaz:2010} and the results in this paper indicate that the instanton partition function is not sensitive to these differences (at least for some theories). For this reason we have not used a nomenclature which emphasises the differences, but this point should be kept in mind in future applications.

Our analysis is far from complete and there are many unsolved problems. It would be desirable to have additional checks (or perhaps even proofs) of the general proposal. One immediate extension is to develop the technology needed to compute conformal blocks for theories with $\cW_3^{(2)}$ symmetry and to compare the results with the proposed dual gauge theory expressions.   The Whittaker vectors and the conformal blocks only depend on the symmetry algebra, but just as for the original AGT conjecture \cite{Alday:2009a} it seems plausible that there is an extension to a relation between correlation functions in the $2d$ CFT and gauge theory partition functions involving some (modified) version of the type studied in \cite{Pestun:2007}. In the general case the relevant  $2d$ CFT is probably a generalised Toda theory (see e.g.~\cite{Feher:1992}), but unfortunately such theories have not been much studied in the literature. 

 Another subject that we did not discuss, but where surface operators appear to be important, is the connections to quantum-mechanical integrable systems. In addition to papers already mentioned this is discussed in e.g.~\cite{Negut:2008}.

\section*{Acknowledgements}

I would like to thank Can Koz\c{c}az,  Sara Pasquetti and Filippo Passerini for collaboration on \cite{Kozcaz:2010b} which formed the basis of the present work. I would also like to thank Nadav Drukker for some useful comments and the string theory group at Queen Mary, University of London for hospitality during the final stages of this work.


\appendix

\setcounter{equation}{0} 
\section{Appendix}

\subsection{The $\sll_3$ Lie algebra} \label{ALie}

The root/weight space of the $\sll_3$  (or $A_{2}$) Lie algebra can be viewed as a $2$--dimensional subspace of $\RR^{3}$. The unit vectors of $\RR^{3}$ are denoted $u_i$ ($i=1,\ldots,3$) and satisfy $\lb u_i,u_j\rb = \de_{ij}$. The simple roots are $e_1 = u_1-u_{2}$ and $e_2=u_2-u_3$.  The positive roots comprise the $e_i$ together with $\tha=e_1+e_2=u_1-u_3$. The fundamental weights are 
\be
\La_1 = \frac{1}{3} ( 2 u_1 -u_2-u_3) \,, \qquad \La_2 = \frac{1}{3} ( u_1+ u_2 - 2u_3)
\ee
and  satisfy $\lb \La_i ,e_j\rb = \de_{ij}$. The Weyl vector, $\rho$, is half the sum of the positive roots, hence $\rho =  \tha = \La_1 + \La_2 $. Finally, the weights of the fundamental representation are 
\be \label{hs}
h_1 = \frac{1}{3} ( 2 u_1 -u_2-u_3)\,, \quad h_2 = \frac{1}{3} ( - u_1 +2u_2-u_3) \,, \quad h_3 = \frac{1}{3} ( - u_1 -u_2+2u_3)
\ee
Note that $h_1 = \La_1$, $h_2=-\La_1+\La_2$ and $h_3 = -\La_2$.

\begingroup\raggedright\endgroup


\begin{thebibliography}{10}

\bibitem{Belavin:1984}
A.~A. Belavin, A.~M. Polyakov, and A.~B. Zamolodchikov, ``{Infinite conformal
  symmetry in two-dimensional quantum field theory},''
{{\em Nucl. Phys.}  {\bfseries B241} (1984) 333--380}.

\bibitem{Nekrasov:2002}
N.~A. Nekrasov, ``{Seiberg-Witten prepotential from instanton counting},'' {\em
  Adv. Theor. Math. Phys.} {\bfseries 7} (2004) 831--864,
{{\ttfamily hep-th/0206161}}.

\bibitem{Alday:2009a}
L.~F. Alday, D.~Gaiotto, and Y.~Tachikawa, ``{Liouville correlation functions
  from four-dimensional gauge theories},''
 {{\em Lett. Math. Phys.}
  {\bfseries 91} (2010) 167--197},
{{\ttfamily arXiv:0906.3219 [hep-th]}}.

\bibitem{Wyllard:2009}
N.~Wyllard, ``{$A_{N-1}$ conformal Toda field theory correlation functions from
  conformal $\cN=2$ $\SU(N)$ quiver gauge theories},''
{{\em JHEP} {\bfseries
  11} (2009) 002},
{{\ttfamily arXiv:0907.2189 [hep-th]}}.

\bibitem{Gaiotto:2009b}
D.~Gaiotto, ``{Asymptotically free $\cN=2$ theories and irregular conformal
  blocks},''
{{\ttfamily arXiv:0908.0307 [hep-th]}}; \\
A.~Marshakov, A.~Mironov, and A.~Morozov, ``{On non-conformal limit of the AGT
  relations},'' {{\em
  Phys. Lett.} {\bfseries B682} (2009) 125--129},
{{\ttfamily arXiv:0909.2052 [hep-th]}}.

\bibitem{Taki:2009}
M.~Taki, ``{On AGT conjecture for pure super Yang-Mills and $\cW$-algebra},''
{{\ttfamily arXiv:0912.4789 [hep-th]}}.

\bibitem{Alday:2010}
L.~F. Alday and Y.~Tachikawa, ``{Affine SL(2) conformal blocks from $4d$ gauge
  theories},''
{{\ttfamily arXiv:1005.4469 [hep-th]}}.

\bibitem{Awata:2010}
H.~Awata, H.~Fuji, H.~Kanno, M.~Manabe, and Y.~Yamada, ``{Localization with a
  surface operator, irregular conformal blocks and open topological string},''
{{\ttfamily arXiv:1008.0574 [hep-th]}}.

\bibitem{Kozcaz:2010b}
C.~Koz\c{c}az, S.~Pasquetti, F.~Passerini, and N.~Wyllard, ``{Affine $\sll(N)$
  conformal blocks from $\cN=2$ $\SU(N)$ gauge theories},''
{{\ttfamily arXiv:1008.1412 [hep-th]}}.

\bibitem{Carlsson:2008}
E.~{Carlsson} and A.~{Okounkov}, ``{Exts and vertex operators},''
{{\ttfamily arXiv:0801.2565 [math.AG]}}.

\bibitem{Braverman:2004a}
A.~Braverman, ``{Instanton counting via affine Lie algebras I: Equivariant
  J-functions of (affine) flag manifolds and Whittaker vectors},''
{{\ttfamily math/0401409}}; \\
A.~Braverman and P.~Etingof, ``{Instanton counting via affine Lie algebras. II:
  From Whittaker vectors to the Seiberg-Witten prepotential},''
{{\ttfamily math/0409441}}.

\bibitem{Yanagida:2010}
S.~Yanagida, ``{Whittaker vectors of the Virasoro algebra in terms of Jack
  symmetric polynomial},''
{{\ttfamily arXiv:1003.1049 [math.QA]}}.

\bibitem{Feigin:2008}
B.~Feigin, M.~Finkelberg, A.~Negut, and L.~Rybnikov, ``{Yangians and cohomology
  rings of Laumon spaces},''
{{\ttfamily arXiv:0812.4656 [math.AG]}}.

\bibitem{Braverman:2010}
A.~Braverman, B.~Feigin, L.~Rybnikov, and M.~Finkelberg, ``{A finite analog of
  the AGT relation I: Finite W-algebras and quasimaps' spaces},''
{{\ttfamily arXiv:1008.3655 [math.AG]}}.

\bibitem{Bershadsky:1989}
A.~A.~Belavin, ``KdV type equations and $\mathcal{W}$ algebras,''
  {\em Advanced Studies in Pure Mathematics} {\bf 19} (1988) 117; \\
M.~Bershadsky and H.~Ooguri, ``{Hidden $\SL(n)$ Symmetry in Conformal Field
  Theories},''
{{\em Commun. Math. Phys.}
  {\bfseries 126} (1989) 49}.

\bibitem{Feigin:1990}
B.~Feigin and E.~Frenkel, ``{Quantization of the Drinfeld-Sokolov reduction},''
{{\em Phys. Lett.}
  {\bfseries B246} (1990) 75--81}; \\
  J.~M.~Figueroa-O'Farrill,
  ``On the homological construction of Casimir algebras,''
 {\em Nucl.\ Phys.}\  {\bf B343} (1990) 450.


\bibitem{Fateev:1987}
V.~A. Fateev and S.~L. Lukyanov, ``{The models of two-dimensional conformal
  quantum field theory with $\ZZ_n$ symmetry},''
{{\em Int. J. Mod. Phys.}
  {\bfseries A3} (1988) 507}.

\bibitem{Bershadsky:1990}
M.~Bershadsky, ``{Conformal field theories via Hamiltonian reduction},''
{{\em Commun. Math. Phys.}
  {\bfseries 139} (1991) 71--82}.

\bibitem{Polyakov:1989}
A.~M. Polyakov, ``{Gauge transformations and diffeomorphisms},''
{{\em Int. J. Mod. Phys.}
  {\bfseries A5} (1990) 833}.

\bibitem{Bais:1990}
F.~A. Bais, T.~Tjin, and P.~van Driel, ``{Covariantly coupled chiral
  algebras},''
{{\em Nucl. Phys.}
  {\bfseries B357} (1991) 632--654}; \\
L.~Feher, L.~O'Raifeartaigh, P.~Ruelle, I.~Tsutsui, and A.~Wipf, ``{Generalized
  Toda theories and $\cW$ algebras associated with integral gradings},''
{{\em Ann. Phys.}
  {\bfseries 213} (1992) 1--20}.

\bibitem{Feher:1992}
L.~Feher, L.~O'Raifeartaigh, P.~Ruelle, I.~Tsutsui, and A.~Wipf, ``{On
  Hamiltonian reductions of the Wess-Zumino-Novikov-Witten theories},''
{{\em Phys. Rept.}
  {\bfseries 222} (1992) 1--64}.

\bibitem{deBoer:1993}
J.~de~Boer and T.~Tjin, ``{The relation between quantum $\cW$ algebras and Lie
  algebras},'' {{\em Commun. Math.
  Phys.} {\bfseries 160} (1994) 317--332}, {{\ttfamily hep-th/9302006}}.

\bibitem{Kac:2003a}
V.~{Kac}, S.~{Roan}, and M.~{Wakimoto}, ``{Quantum reduction for affine
  superalgebras},'' {{\em
  Commun. Math. Phys.} {\bfseries 241} (2003) 307--342},
{{\ttfamily math-ph/0302015}}.

\bibitem{Gukov:2006}
S.~Gukov and E.~Witten, ``{Gauge theory, ramification, and the geometric
  langlands program},''
{{\ttfamily hep-th/0612073}}.

\bibitem{Gukov:2007}
S.~Gukov, ``{Surface operators and knot homologies},''
{{\ttfamily arXiv:0706.2369 [hep-th]}}; \\
M.-C. Tan, ``{Supersymmetric surface operators, four-manifold theory and
  invariants in various dimensions},''
{{\ttfamily arXiv:1006.3313 [hep-th]}}.

\bibitem{Alday:2009b}
L.~F. Alday, D.~Gaiotto, S.~Gukov, Y.~Tachikawa, and H.~Verlinde, ``{Loop and
  surface operators in $\cN=2$ gauge theory and Liouville modular geometry},''
{{\em JHEP} {\bfseries 01}
  (2010) 113},
{{\ttfamily arXiv:0909.0945 [hep-th]}}.

\bibitem{Gaiotto:2009c}
D.~Gaiotto, ``{Surface operators in $\cN=2$ $4d$ gauge theories},''
{{\ttfamily arXiv:0911.1316 [hep-th]}}.

\bibitem{Kozcaz:2010}
C.~Koz\c{c}az, S.~Pasquetti, and N.~Wyllard, ``{A $\&$ B model approaches to
  surface operators and Toda theories},'' JHEP {\bf 1008} (2010) 042
, {{\ttfamily arXiv:1004.2025 [hep-th]}}.

\bibitem{Dimofte:2010}
T.~Dimofte, S.~Gukov, and L.~Hollands, ``{Vortex counting and Lagrangian
  3-manifolds},''
{{\ttfamily arXiv:1006.0977 [hep-th]}}.

\bibitem{Maruyoshi:2010}
K.~Maruyoshi and M.~Taki, ``{Deformed prepotential, quantum integrable system
  and Liouville field theory},''
{{\ttfamily arXiv:1006.4505 [hep-th]}}.

\bibitem{Taki:2010}
M.~Taki, ``{Surface operator, bubbling Calabi-Yau and AGT relation},''
{{\ttfamily arXiv:1007.2524 [hep-th]}}.

\bibitem{Zamolodchikov:1985}
A.~B. Zamolodchikov, ``{Infinite additional symmetries in two-dimensional
  conformal quantum field theory},''
{{\em Theor. Math. Phys.} {\bfseries
  65} (1985) 1205--1213}.

\bibitem{Drukker:2010}
N.~Drukker, D.~Gaiotto, and J.~Gomis, ``{The virtue of defects in $4D$ gauge
  theories and $2D$ CFTs},''
{{\ttfamily arXiv:1003.1112 [hep-th]}}.

\bibitem{Ademollo:1975}
M.~Ademollo {\em et al.}, ``{Supersymmetric strings and color confinement},''
{{\em Phys. Lett.}
  {\bfseries B62} (1976) 105}.

\bibitem{Tjin:1992}
T.~Tjin, ``{Finite $\cW$ algebras},''
{{\em Phys. Lett.}
  {\bfseries B292} (1992) 60--66},
{{\ttfamily hep-th/9203077}}.

\bibitem{deBoer:1992}
J.~de~Boer and T.~Tjin, ``{Quantization and representation theory of finite $\cW$
  algebras},'' {{\em Commun. Math.
  Phys.} {\bfseries 158} (1993) 485--516},
{{\ttfamily hep-th/9211109}}.

\bibitem{DeSole:2005}
A.~{De Sole} and V.~{Kac}, ``{Finite vs. affine $\cW$-algebras},''
 {{\ttfamily math-ph/0511055}}.

\bibitem{Kac:2004}
V.~G. {Kac} and M.~{Wakimoto}, ``{Quantum reduction in the twisted case},''
{{\ttfamily math-ph/0404049}}; \\
B.~Noyvert, ``{Ramond sector of superconformal algebras via quantum
  reduction},'' {\em JHEP} {\bfseries 11} (2006) 045,
{{\ttfamily math-ph/0408061}}; \\
T.~{Arakawa}, ``{Representation theory of $\cW$-algebras, II: Ramond twisted
  representations},'' {{\ttfamily
  arXiv:0802.1564 [math.QA]}}.

\bibitem{Smith:1990}
S.~P. Smith, ``{A class of algebras similar to the enveloping algebra of $
  \mathfrak{sl}_2$},'' {\em Trans. Amer. Math. Soc.} {\bfseries 322} (1990)
  285.

\bibitem{Furlan:1994}
P.~Furlan, A.~C. Ganchev, and V.~B. Petkova, ``{Singular vectors of $\cW$ algebras
  via DS reduction of $A_2^{(1)}$},''
 {{\em Nucl. Phys.}
  {\bfseries B431} (1994) 622--666},
{{\ttfamily hep-th/9403075}}.

\bibitem{Mironov:2009a}
A.~Mironov and A.~Morozov, ``{The power of Nekrasov functions},''
 {{\em Phys. Lett.}
  {\bfseries B680} (2009) 188--194},
{{\ttfamily arXiv:0908.2190 [hep-th]}}; \\
A.~Mironov and A.~Morozov, ``{On AGT relation in the case of U(3)},''
  {{\em Nucl. Phys.}
  {\bfseries B825} (2010) 1--37},
{{\ttfamily arXiv:0908.2569 [hep-th]}}.

\bibitem{Kanno:2009}
  S.~Kanno, Y.~Matsuo, S.~Shiba and Y.~Tachikawa,
  ``$\cN=2$ gauge theories and degenerate fields of Toda theory,''
  {\em Phys.\ Rev. } {\bf D81}, 046004 (2010)
  {\ttfamily arXiv:0911.4787 [hep-th]};  \\
 S.~Kanno, Y.~Matsuo and S.~Shiba,
  ``Analysis of correlation functions in Toda theory and AGT-W relation for
  $\SU(3)$ quiver,''
  {\em Phys.\ Rev.} {\bf D82}, 066009 (2010)
  {\ttfamily arXiv:1007.0601 [hep-th]}; \\
   F.~Passerini,
  ``Gauge theory Wilson loops and conformal Toda field theory,''
  JHEP {\bf 1003} (2010) 125,  
  {\tt arXiv:1003.1151 [hep-th]}; \\
 J.~Gomis and B.~Le Floch,
  ``'t Hooft operators in gauge theory from Toda CFT,''
  {\ttfamily arXiv:1008.4139 [hep-th]}.

\bibitem{Fateev:2005}
V.~A. Fateev and A.~V. Litvinov, ``{On differential equation on four-point
  correlation function in the conformal Toda field theory},''
 {{\em JETP Lett.} {\bfseries 81}
  (2005) 594--598},
{{\ttfamily hep-th/0505120}}; \\
V.~A. Fateev and A.~V. Litvinov, ``{Correlation functions in conformal Toda
  field theory I},''
 {{\em JHEP} {\bfseries
  11} (2007) 002},
{{\ttfamily arXiv:0709.3806 [hep-th]}}.

\bibitem{Schiappa:2009}
R.~Schiappa and N.~Wyllard, ``{An $A_r$ threesome: Matrix models, 2d CFTs and
  4d N=2 gauge theories},''
{{\ttfamily arXiv:0911.5337 [hep-th]}}.

\bibitem{Dijkgraaf:2009}
R.~Dijkgraaf and C.~Vafa, ``{Toda theories, matrix models, topological strings,
  and $\cN=2$ gauge systems},''
{{\ttfamily arXiv:0909.2453 [hep-th]}}.

\bibitem{Cheng:2010}
M.~C.~N. Cheng, R.~Dijkgraaf, and C.~Vafa, ``{Non-perturbative topological
  strings and conformal blocks},''
{{\ttfamily arXiv:1010.4573 [hep-th]}}.

\bibitem{Awata:2009}
H.~Awata and Y.~Yamada, ``{Five-dimensional AGT conjecture and the deformed
  Virasoro algebra},'' {{\em
  JHEP} {\bfseries 01} (2010) 125},
{{\ttfamily arXiv:0910.4431 [hep-th]}}.

\bibitem{Gaiotto:2009a}
D.~Gaiotto, ``{$\cN=2$ dualities},''
{{\ttfamily arXiv:0904.2715 [hep-th]}}.

\bibitem{Pestun:2007}
V.~Pestun, ``{Localization of gauge theory on a four-sphere and supersymmetric
  Wilson loops},''
{{\ttfamily arXiv:0712.2824 [hep-th]}}.

\bibitem{Negut:2008}
A.~{Negut}, ``{Laumon spaces and the Calogero-Sutherland integrable system},''
  {\em Invent. Math.} {\bfseries 178} (2009) 299,
{{\ttfamily arXiv:0811.4454 [math.AG]}}; \\
N.~A. Nekrasov and S.~L. Shatashvili, ``{Quantization of Integrable Systems and
  Four Dimensional Gauge Theories},''
{{\ttfamily arXiv:0908.4052 [hep-th]}}; \\
A.~Mironov and A.~Morozov, ``{Nekrasov functions from exact BS periods: the
  case of $\SU(N)$},'' {\em J. Phys.} {\bfseries A43} (2010) 195401,
{{\ttfamily arXiv:0911.2396 [hep-th]}}; \\
N.~Nekrasov and E.~Witten, ``{The Omega Deformation, Branes, Integrability, and
  Liouville Theory},''
{{\ttfamily arXiv:1002.0888 [hep-th]}}; \\
J.~Teschner, ``{Quantization of the Hitchin moduli spaces, Liouville theory,
  and the geometric Langlands correspondence I},''
{{\ttfamily arXiv:1005.2846 [hep-th]}}.

\end{thebibliography}
\end{document}